\documentclass{emulateapj}
\shorttitle{Supernova shock breakout light curves}
\shortauthors{SUZUKI\&SHIGEYAMA}
\begin{document}
\title{PROBING EXPLOSION GEOMETRY OF CORE-COLLAPSE SUPERNOVAE WITH LIGHT CURVES OF THE SHOCK BREAKOUT}
\author{AKIHIRO SUZUKI\altaffilmark{1,2} and TOSHIKAZU SHIGEYAMA\altaffilmark{1}}
\altaffiltext{1}{Research Center for the Early Universe, School of Science, University of Tokyo, Bunkyo-ku, Tokyo 113-0033, Japan.}
\altaffiltext{2}{Department of Astronomy, School of Science, University of Tokyo, Bunkyo-ku, Tokyo 113-0033, Japan.}
\begin{abstract}
We investigate effects of aspherical energy deposition in core-collapse supernovae on the light curve of the supernova shock breakout. 
We performed two-dimensional hydrodynamical calculations of an aspherical supernova explosion to obtain the time when a shock wave generated in the stellar interior reaches the stellar surface in each radial direction. 
Using results of the calculations, light curves during the shock breakout are derived in an approximate way. 
We show that the light curve during the shock breakout can be a strong indicator of aspherical properties of core-collapse supernovae. 
\end{abstract}
\section{INTRODUCTION\label{intro}}
Massive stars with their main-sequence mass greater than 8$M_\odot$ end their lives as core-collapse supernovae (CCSNe). 
The very beginning of dynamical evolution of a CCSN, at which a strong shock wave resulting from the collapse of the iron core abruptly emerges from the stellar atmosphere, is called the supernova shock breakout. 
Earlier theoretical investigations \citep{kc78,f78} predicted that UV/soft X-ray photons having been produced in the stellar interior begin to escape from the stellar surface at the moment of the shock emergence. 
The enormous amount of the escaping photons makes the moment the brightest phase throughout the evolution of a CCSN. 
Although observations of the shock breakout are very challenging due to its short duration (ranging from a few to a thousand seconds), some cases of the detection of the shock breakout have been reported, SN 2008D \citep{s08,l08,m08,m09}, SNLS-04D2dc \citep{sc08,g08}, and SNLS-06D1jd \citep{g08}. 
The growing examples offer us good opportunities for comparing the theory of the shock breakout with observations.

The nature of the shock breakout has been investigated intensively in the framework of semi-analytical approaches based on self-similar solutions \citep{in88,mm99,ss10a} and numerical simulations \citep{s88,w88,bn91,eb92,to09}. 
For the proper treatment of  the coupling of radiation and matter, radiation-hydrodynamics codes are often used. 
\cite{s88} performed radiation-hydrodynamical calculations based on the flux-limited diffusion approximation. 
\cite{bn91} and \cite{eb92} developed multi-group radiation-hydrodynamics codes for the modeling of the phenomenon. 
Although there are differences between methods to model the phenomenon, these works reached the following general agreements: i) the temperature behind the shock wave reaches about $10^5$ - $10^6$ K, ii) the matter behind the shock wave emits UV/soft X-ray photons having a black body spectrum, iii) the light travel time across the progenitor star is crucial to determine the light curves of the emissions. 
It is noteworthy that the observed spectrum of the shock breakout of SN 2008D is well fitted by a power-law \citep{s08,l08,m08,m09}, which contradicts the theoretical expectation. 
Recent investigations \citep{wwm07,ss10a} suggest that interactions between thermal photons and electrons at the shock front via Compton scattering, which is so-called bulk comptonization \citep{bp81a,bp81b}, can form a power-law spectrum. 

These investigations assume spherical symmetry, i.e., the shock breakout is assumed to occur simultaneously at every point on the surface of the progenitor star. 
However, recent studies on the explosion mechanism of CCSNe strongly indicate aspherical deposition of the explosion energy \citep{m99,b03,k04}, although details of the energy deposition process remain unclear. 
From the observational viewpoint, CCSNe are considered to be aspherical in general \citep[e.g.,][]{ma08}. 
For an aspherical explosion, the shock breakout does not occur simultaneously in different radial directions. 
As a result, the light curve during the shock breakout of an aspherical explosion can deviate from those of spherical one. 
In other words, the light curve during the shock breakout can be used as a probe for the explosion geometry of a CCSN. 
\cite{c09} performed a series of hydrodynamical simulations of jet-induced explosions of a red supergiant progenitor. 
They investigated only the difference of light curves during the shock breakout between thermal and kinetic energy-dominated jets. 
However, light curves during the shock breakout must contain more information on the explosion geometry of a CCSN, e.g., the viewing angle, the degree of aspherical energy deposition, and so on. 

In this letter, we present an approximate method to calculate light curves during the shock breakout assuming axisymmetric energy deposition and show that the thus calculated light curves indeed reflect the explosion geometry of a CCSN. 
In \S 2, we describe the procedure to calculate light curves during the shock breakout. 
Resultant light curves are shown in \S 3. 
Finally, \S 4 concludes this paper.  

\section{FORMULATION\label{formulation}}
\begin{figure}[t]
\begin{center}
\includegraphics[scale=0.55]{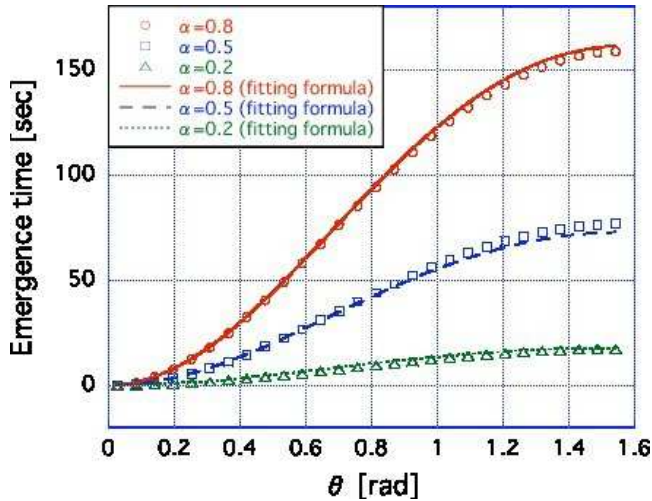}
\caption{The emergence time as a function of the angle $\theta$ for various $\alpha$. 
The circles, squares and triangles are correspond to  models with $\alpha=0.8, 0.5$, and $0.2$, respectively. 
The solid, dashed, and dotted curves are calculated by the empirical formula (\ref{empirical}) for $\alpha=0.8, 0.5$, and $0.2$. 
}
\label{emergence}
\end{center}
\end{figure}
\begin{figure}[b]
\begin{center}
\includegraphics[scale=0.4]{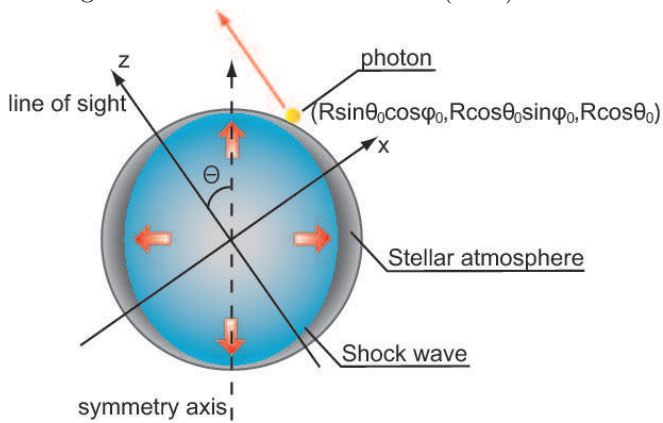}
\caption{A schematic view of the geometry of the aspherical shock breakout.}
\label{schematic}
\end{center}
\end{figure}

In this section, we describe our hydrodynamical models and the procedure to calculate light curves during the shock breakout. 
\subsection{Hydrodynamical model}
As we noted in the previous section, the time when the shock wave propagating in different radial directions reaches the stellar surface (referred to as the emergence time in this paper) can take various values.  
To calculate light curves during the shock breakout, we must know the emergence time at each point on the stellar surface. 
Then, we performed a series of hydrodynamical calculations of a supernova explosion of a blue supergiant progenitor using a two-dimensional relativistic Eulerian hydrodynamics code, which is developed according to \cite{db02}. 
Details of the calculations are described in the forthcoming paper \citep{ss10b}. 
In the following, we present the brief summary of the calculations. 
The code integrates the basic equations of the special relativistic hydrodynamics in two-dimensional spherical polar coordinates $(r,\theta)$, where $r$ is the distance from the origin and $\theta$ is the angle measured from the symmetry axis. 
The equation of state for an ideal gas of the adiabatic index $4/3$ is assumed. 
The progenitor model is the same as used in \cite{s90}, i.e., a progenitor model of SN 1987A. 
To generate a strong shock wave propagating in the stellar interior, we impose the following initial condition on the radial component $v_\mathrm{in}$ of the velocity at the inner boundary ($r=r_\mathrm{in}\simeq 10^8$ cm),
\begin{equation}
v_\mathrm{in}(\theta)=\sqrt{\frac{2E_\mathrm{exp}}{4\pi r_\mathrm{in}^2\Delta r_\mathrm{in}\rho_\mathrm{in}}}
\frac{1+\alpha\cos(2\theta)}{\sqrt{1-2\alpha/3+7\alpha^2/15}},
\label{initial}
\end{equation}
where $E_\mathrm{exp}=10^{51}$ erg is the explosion energy, $\rho_\mathrm{in}$ is the density at the inner boundary, and $\Delta r_\mathrm{in}$ is the width of the radial zone at the inner boundary. 
Here we have introduced a parameter $\alpha$ characterizing the extent of the aspherical explosion. 
Using this parameter, we can express the ratio of the radial component of the velocity at $(r_\mathrm{in},0)$ to that at $(r_\mathrm{in},\pi/2)$ as $v_\mathrm{in}(0)/v_\mathrm{in}(\pi/2)=(1+\alpha)/(1-\alpha)$. 
\cite{m02} calculated the density and the velocity distribution of ejected matter as a result of aspherical supernova explosions by a series of two-dimensional hydrodynamical calculations and obtained the distribution of heavy elements, such as $^{56}$Ni, synthesized in the ejecta. 
\cite{m06} studied the optical emission from the ejecta by using a Monte-Carlo radiative transfer calculation. 
They found that the axis ratio of the Ni blob in the ejecta is at most a few even for the model with $v_\mathrm{in}(0)/v_\mathrm{in}(\pi/2)=8$, which corresponds to $\alpha=7/9$ in our model. 
Therefore, all models considered here must be observed as explosions with the axis ratio smaller than a few. 
It should be noted that there are many other candidates for the form of the angular dependence of the initial condition for $v_\mathrm{in}$. 
In this study, we adopt one of the simplest form, eq. (\ref{initial}), because it has an advantage in clarifying general behaviors of light curves during the shock breakout originated from an aspherical explosion.  
However, due to this simple treatment of asphericity, our model fails to reproduce gamma-ray burst associated SNe even if we choose large $\alpha$. 
Results of simulations of jet-driven explosions \citep[e.g.,][]{tom09} show that matter located near the equatorial plane fall back to the central object, which strongly affects the velocity of the shock wave propagating along the direction. 
To calculate the propagation velocity of the shock accurately, we need to perform simulations of the explosion including self-gravity. 
For the boundary condition at the stellar surface, we assume the free boundary condition, i.e., the derivatives of physical quantities with respect to $r$ at the stellar surface are assumed to be zero.

The thus generated shock wave propagates outward and finally reaches the surface. 
Figure \ref{emergence} shows the emergence time $T(\alpha,\theta)$ as a function of the angle $\theta$ for various $\alpha$. 
The emergence time is defined so that $T(\alpha,0)=0$. 
The circles, squares and triangles in Figure \ref{emergence} correspond to  models with $\alpha=0.8, 0.5$, and $0.2$, respectively. 
We find that the behavior of the emergence time in the range of $0\le\alpha\le0.8$ is well approximated by the following empirical formula,
\begin{equation}
T(\alpha,\theta)=65\alpha(1+3.6\alpha)(\sin^2\theta-0.2\sin^4\theta).
\label{empirical}
\end{equation}
The lines in Figure \ref{emergence} are plotted using this formula. 

To model the shock breakout accurately, we must treat the coupling of radiation and matter by solving the equations of radiation-hydrodynamics, because the optical depth of the shock falls below unity after the shock emergence. 
However, thermal photons having been emitted by the shock front diffuse out from the envelope when the optical depth $\tau_\mathrm{s}$ of the shock decreases to the speed of light divided by the shock velocity, $c/V_\mathrm{s}$. 
In the case of the explosion of a blue supergiant progenitor with the explosion energy of $E_\mathrm{exp}=10^{51}$ erg, the shock velocity reaches $\sim 0.03c$ at the moment of the shock emergence, which leads to the optical depth at the shock breakout of $\tau_\mathrm{s}\sim 30$. 
The radiation field at such a large optical depth is expected to approach the black body. 
Furthermore, in the Rosseland approximation, the fraction of the anisotropic part of the radiation field is of the order of  $4/(3\pi\tau_\mathrm{s})$, which is much smaller than unity. 
Thus, we assume that isotropic black body radiation with a constant flux lasts for a time interval $\tau$ after the shock reaches the stellar surface. 
We need the temperature $T_\mathrm{r}$ of the radiation emitted by the matter behind the shock wave when the shock reaches the surface in order to evaluate the flux at each point on the stellar surface. 
In the hydrodynamical calculations, we assumed the equation of state for an ideal gas of the adiabatic index $4/3$, which is a good approximation if the pressure is radiation dominated. 
Therefore, we estimate $T_\mathrm{r}$ by assuming that the post-shock pressure $p_\mathrm{s}$ at the outer boundary is equal to the radiation pressure given by $aT_\mathrm{r}^4/3$, where $a(=7.566\times 10^{-15}$ erg cm$^{-3}$ K$^{-4}$) is the radiation constant. 
We find that the post-shock pressure when the shock reaches the stellar surface is almost independent of the angle $\theta$, $p_\mathrm{s}\simeq 2.4\times 10^9$ g/cm s$^2$, which leads to the radiation temperature of $T_\mathrm{r}\simeq 10^6$ K.  

\subsection{Derivation of light curves}
We calculate light curves during the shock breakout for a given emergence time distribution $T(\alpha,\theta)$. 
A schematic view of the geometry considered here is shown in Figure \ref{schematic}. 

At first, we introduce the viewing angle $\Theta$ as the angle between the line of sight and the symmetry axis. 
We define cartesian coordinates $(x,y,z)$ so that the $z$-axis is identical with the line of sight. 
The symmetry axis is assumed to be in the $x$-$z$ plane. 
Thus, a unit vector parallel to the symmetry axis is expressed as $(\sin\Theta,0,\cos\Theta)$ in the coordinate system. 
We consider the flux $F(t,\theta_0,\phi_0,\alpha)$ at a point $(R\sin\theta_0\cos\phi_0,R\sin\theta_0\sin\phi_0,R\cos\theta_0)$ on the surface of a progenitor star with the radius $R(\simeq 3\times 10^{12}\ \mathrm{cm})$ at a time $t$. 
Using the angle $\chi$ between the symmetry axis and the position vector of the point given by
\begin{equation}
\cos\chi=\cos\theta_0\cos\Theta+\sin\theta_0\sin\Theta\cos\phi_0,
\label{chi}
\end{equation}
we can express the emergence time at the point as $T(\alpha,\chi)$. 
As we noted in the previous section, we assume that black body radiation with a constant flux lasts for a time interval $\tau$ after the shock emergence. 
The time interval $\tau$ is the diffusion time scale of photons emitted from the shock front. 
The value depends on the explosion energy, the mass of the ejected matter, and the structure of the envelope. 
\cite{mm99} studied the shock breakout emission by using the self-similar solution derived by \cite{s60} and confirmed that the results are consistent with those of numerical simulations. 
They derived $40$ s for the explosion of a blue supergiant progenitor with the explosion energy of $10^{51}$ ergs and the ejecta mass of $10M_\odot$.  
Thus, we set $\tau=40$ s in this study.  
The flux is expressed as
\begin{equation}
F(t,\theta_0,\phi_0,\alpha)=\left\{\begin{array}{ccl}
\sigma_\mathrm{SB}T_\mathrm{r}^4&\mathrm{for}&T(\alpha,\chi)<t<T(\alpha,\chi)+\tau,\\
0&&\mathrm{otherwise},
\end{array}\right.
\label{emissivity}
\end{equation}
where $\sigma_\mathrm{SB}=5.67\times10^{-5}$ erg cm$^{-2}$ s$^{-1}$ K$^{-4}$ is the Stefan-Boltzmann constant. 

Next, we consider an observer located at $(0,0,D)$. 
The distance $D$ is sufficiently larger than the radius $R$, i.e., the progenitor star is regarded as a point source for the observer. 
A photon observed at a time $t$ was emitted from the source position $(R\sin\theta_0\cos\phi_0,R\sin\theta_0\sin\phi_0,R\cos\theta_0)$ at $t_0=t-(D-R\cos\theta_0)/c$. 
Therefore, the luminosity $L(t,\alpha,\Theta)$ at the position of the observer at $t$ is obtained by integrating the flux $F(t_0,\theta_0,\phi_0)$ multiplied by $R^2\sin\theta_0$ with respect to $0\le\theta_0<\pi/2$ and $0\le\phi_0<2\pi$,
\begin{equation}
L(t,\alpha,\Theta)=R^2\int^{\pi/2}_0\int^{2\pi}_0F(t_0,\theta_0,\phi_0,\alpha)\sin\theta_0d\theta_0d\phi_0.
\label{power}
\end{equation}

When we approximate the form of the emergence time as $T(\alpha,\theta)=A(\alpha) \sin^2\theta$ and assume $\Theta=0$, we can carry out the integration in the above formula (\ref{power}) and derive the following approximate formulae of the luminosity,
\begin{equation}
L(t,\alpha,0)\simeq\left\{
\begin{array}{l}
2\pi R^2\sigma_\mathrm{SB}T_\mathrm{r}^4\left[1-f(\alpha,t)\right]\\
\hspace{3em}\mathrm{for}\ \ \frac{D-R}{c}<t<\frac{D-R}{c}+\tau,\\
2\pi R^2\sigma_\mathrm{SB}T_\mathrm{r}^4\left[f(\alpha,t-\tau)-f(\alpha,t)\right]\\
\hspace{3em}\mathrm{for}\ \ \frac{D-R}{c}+\tau<t<\frac{D}{c}+A(\alpha),\\
2\pi R^2\sigma_\mathrm{SB}T_\mathrm{r}^4f(\alpha,t-\tau)\\
\hspace{3em}\mathrm{for}\ \ \frac{D}{c}+A(\alpha)<t<\frac{D}{c}+\tau+A(\alpha),\\
0\ \ \ \ \ \mathrm{otherwise},\end{array}\right.
\end{equation}
where we have defined a function $f(\alpha,t)$ as
\begin{equation}
f(\alpha,t)=\frac{-R+\sqrt{R^2-4cA(\alpha)[ct-cA(\alpha)-D]}}{2cA(\alpha)}.
\end{equation}
We find that the above approximate formula agrees with numerically calculated light curves within $~$10 percent error if we define the function $A(\alpha)$ as
\begin{equation}
A(\alpha)=52\alpha(1+3.6\alpha). 
\end{equation}

\begin{figure}[t]
\begin{center}
\includegraphics[scale=0.55]{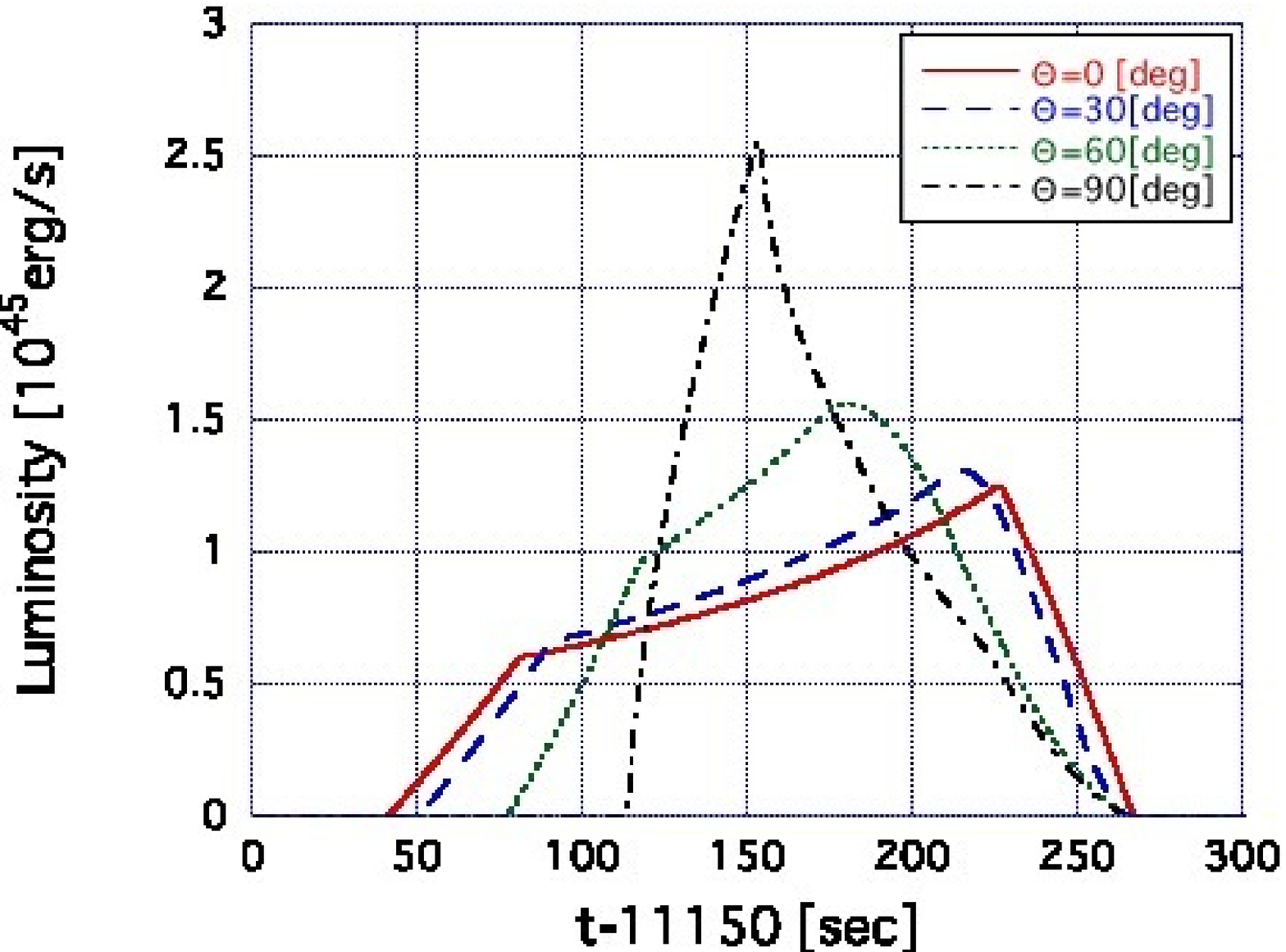}
\caption{Light curves during the shock breakout for various viewing angles $\Theta$ and a fixed $\alpha(=0.5)$. 
The solid, dashed, dotted and dash-dotted lines represent models with $\Theta=0\degr, 30\degr, 60\degr$, and $90\degr$, respectively.}
\label{view}
\end{center}
\end{figure}

\begin{figure}[t]
\begin{center}
\includegraphics[scale=0.55]{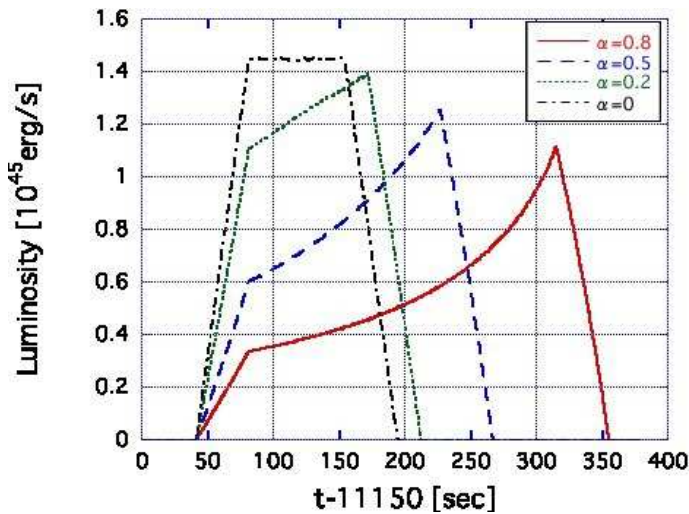}
\caption{Light curves during the shock breakout for a fixed viewing angle $\Theta=0\degr$ and various $\alpha$. 
The solid, dashed, dotted, and dash-dotted lines represent models with $\alpha=0.8, 0.5, 0.2$, and $0$, respectively.}
\label{f4}
\end{center}
\end{figure}

\begin{figure}[t]
\begin{center}
\includegraphics[scale=0.55]{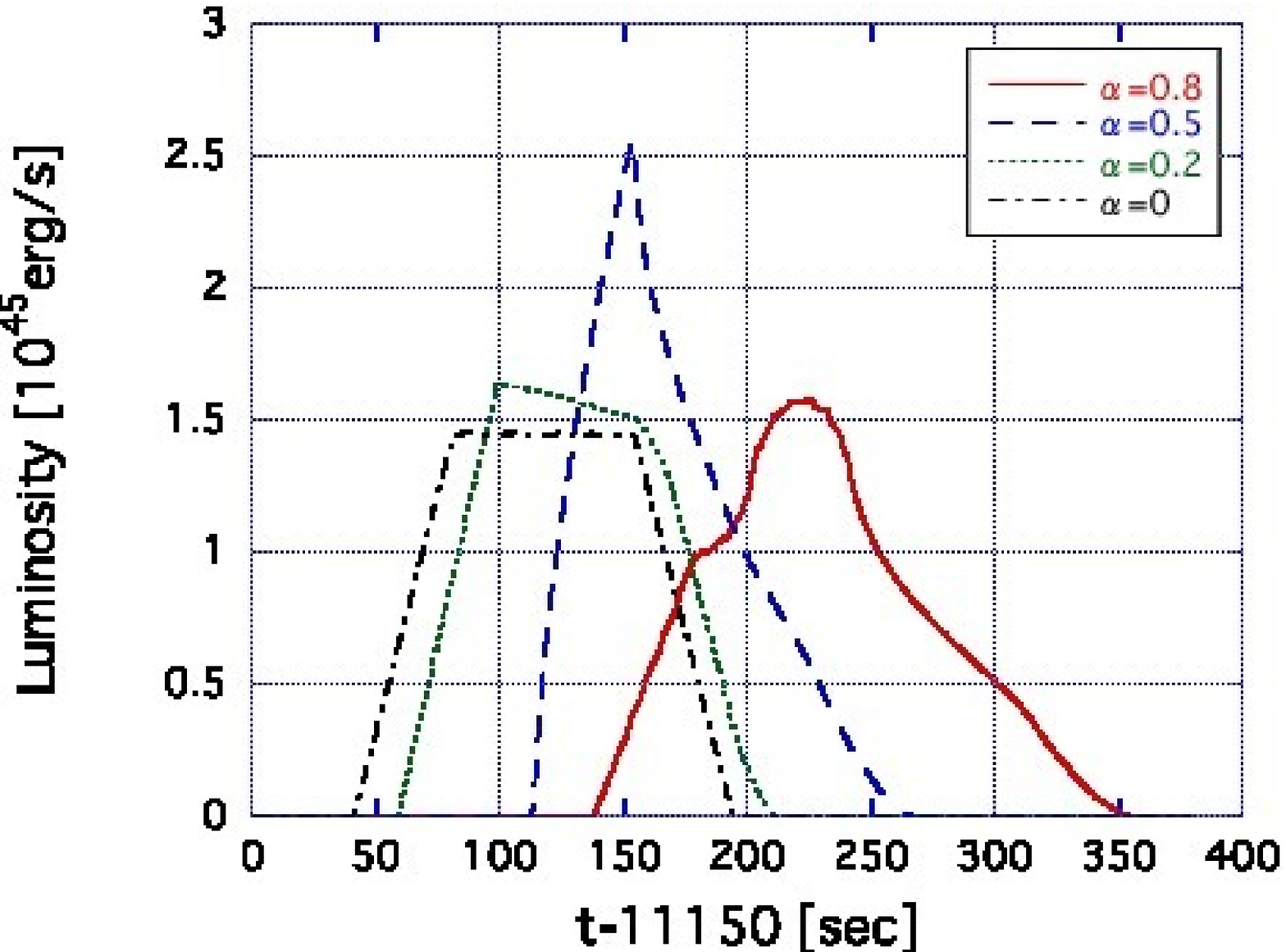}
\caption{Same as Figure \ref{f4}, but for $\Theta=90\degr$.}
\label{f5}
\end{center}
\end{figure}

\section{RESULTS\label{results}}
We calculate light curves during the shock breakout by using eqs. (\ref{empirical}), (\ref{chi}), (\ref{emissivity}), and (\ref{power}). 
The distance $D$ is set to be $D=100R$.  

\cite{eb92}, who tackled the problem with a multi-group radiation-hydrodynamics code, calculated light curves of the breakout emission for a similar progenitor model. 
Some general features of their results are: i) the peak luminosity reaches $\sim 10^{45}$ erg/s, ii) the duration of the emission is a few minutes, iii) the radiation temperature is of the order of $10^6$ K. 
Despite the approximate treatment of emissions from the shock, light curves calculated by our method successfully reproduce the features. 

\subsection{The dependence on the viewing angle}
Figure \ref{view} shows light curves for various viewing angles $\Theta$ and a fixed $\alpha(=0.5)$. 
A clear distinction between models with small viewing angles and those with large viewing angles is recognized. 
For small viewing angle models, the luminosity gradually rises after a sudden brightening and eventually reaches the peak value. 
For large viewing angle models, on the other hand, the luminosity suddenly reaches the peak value and then begins to decline. 

This behavior is easily explained by considering how photons emitted from the stellar surface travel to the observer. 
In the case of small viewing angles, the shock wave initially emerges from the point on the symmetry axis ($\theta_0\simeq0$) and then points with $\theta_0>0$ experience the shock emergence one after another. 
Because the emergence time distributions $T(\alpha,\theta)$ flattens at $\theta=\pi/2$ as shown in Figure \ref{emergence}, the shock emergence at points with $\theta_0\simeq \pi/2$ is expected to occur almost simultaneously. 
In other words, in a later epoch, the observer receives more photons than in the earlier epoch. 
This is the reason why the luminosity reaches the peak value after a gradual increase for small viewing angle. 

Next, we consider models with large viewing angles. 
In this case, the shock wave initially emerges from points with $\theta_0\simeq\pi/2$. 
Photons emitted from the points need longer time to reach the position of the observer than those from points with $\theta_0<\pi/2$. 
Because of the combined effect of the time delay and the early emergence of the shock, photons from points with $\theta_0\simeq \pi/2$ reach the position of the observer almost simultaneously, which leads to a more sudden brightening to the peak luminosity. 

\subsection{The dependence on the parameter $\alpha$}
Next, we investigate how the degree of asphericity of an explosion, which is characterized by the parameter $\alpha$ in eq. (\ref{initial}), affects the light curve during the shock breakout. 

Figure \ref{f4} shows light curves for a fixed viewing angle $\Theta(=0\degr)$ and various $\alpha$. 
The dash-dotted line represents the model with $\alpha=0$, i.e., a spherical explosion. 
In this case, the shock breakout simultaneously occurs in the same way at every points on the stellar surface. 
Therefore, the luminosity does not evolve with time. 
The duration of the emission is roughly determined by light-travel-time across the stellar radius, $R/c\simeq 100$ s. 
For models with a larger $\alpha$, the duration of the emission is longer than that of the model with $\alpha=0$. 
This is because the shock emerges at points with $\theta_0>0$ after it does at the point with $\theta_0=0$. 
Therefore, the duration of the emission is roughly estimated as $R/c+T(\alpha,\pi/2)$. 
Furthermore, since the total energy of the burst is same for all models, the average luminosity is small for models with large $\alpha$.

Figure \ref{f5} shows light curves for a fixed viewing angle $\Theta(=90\degr)$ and various $\alpha$. 
In this case, the duration of the emission takes a similar value for all models. 
For models with a large $\alpha$, the feature claimed above, i.e., the sudden rise and the subsequent gradual decrease of the luminosity, is prominent.

\section{CONCLUSIONS\label{conclusions}}
In this letter, we investigate effects of an aspherical explosion on the light curves during the supernova shock breakout. 
To obtain the emergence time distribution $T(\alpha,\theta)$, we performed a series of two-dimensional hydrodynamical calculations of aspherical supernova explosions. 
We establish an approximate way to calculate light curves during the shock breakout using results of the simulations and confirm that resultant light curves successfully reproduce some features of the emission from the supernova shock breakout. 

We find that the light curves strongly depend on the viewing angle and the degree of asphericity of an explosion. 
For small viewing angles from the symmetry axis, the luminosity gradually increases and then reaches the peak value. 
On the contrary, for large viewing angles, the luminosity suddenly reaches the peak value and then gradually decreases. 
For a larger degree of asphericity, the light curve can be distinguished from that of a spherical symmetric explosion irrespective of the viewing angle. 
These features allow us to use light curves during the supernova shock breakout as a probe for the explosion geometry of CCSNe. 
It is noteworthy that the light curve of the shock breakout of SN 2008D \citep{s08} is similar to that of models with large viewing angles, e.g., the model with $\alpha=0.5$ and $\Theta=90$, except for the duration. 
The nebular phase observations of SN 2008D \citep{t09} revealed that SN 2008D was a side-viewed bipolar explosion, which also prefers large viewing angle models.

In this study, we only calculate the supernova shock breakout from the progenitor of SN 1987A, i.e., a blue supergiant, in order to confirm the validity of our approximate way to calculate light curves by comparing results of other studies that adopt more sophisticated ways. 
However, since the duration of the shock breakout from red supergiant progenitors, whose radii are typically of the order of $10^{13}$ cm, is expected to be longer than those from blue supergiant progenitors \citep{mm99}, the shock breakout from a red supergiant progenitor is likely to be observed by future observations. 
In the forthcoming paper \citep{ss10b}, we will present results of the calculations for other progenitor models as well as the description of details of the hydrodynamical simulations.

\acknowledgments
This work has been partly supported by  Grant-in-Aid for JSPS Fellows (21$\cdot$1726) and Scientific Research on Priority Areas (21018004) of the
Ministry of Education, Culture, Sports, Science, and Technology in Japan.

\end{document}